\documentclass[pra,preprint,superscriptaddress,showpacs]{revtex4-1}

\usepackage{graphicx}
\usepackage{bm}                 
\usepackage{amsmath}            
\usepackage{amssymb}
\usepackage{verbatim}           
\usepackage{amsthm}             
\theoremstyle{plain}            

\def\bra#1{{\langle#1|}}
\def\ket#1{{|#1\rangle}}

\def\tr{{\rm Tr}}

\begin{document}

\title{Quantum field theory from a quantum cellular automaton in one spatial dimension and a no-go theorem in higher dimensions}

\author{Leonard \surname{Mlodinow}}
\author{Todd A. \surname{Brun}}\email{tbrun@usc.edu}
\affiliation{Center for Quantum Information Science and Technology, University of Southern California, Los Angeles, California}

\date{\today}

\begin{abstract}
It has been shown that certain quantum walks give rise to relativistic wave equations, such as the Dirac and Weyl equations, in their long-wavelength limits.  This intriguing result raises the question of whether something similar can happen in the multi-particle case.  We construct a one-dimensional quantum cellular automaton (QCA) model which matches the quantum walk in the single particle case, and which approaches the quantum field theory of free fermions in the long-wavelength limit.  However, we show that this class of constructions does not generalize to higher spatial dimensions in any straightforward way, and that no construction with similar properties is possible in two or more spatial dimensions.  This rules out the most common approaches based on QCAs.  We suggest possible methods to overcome this barrier while retaining locality.
\end{abstract}

\pacs{}

\maketitle

\section{Introduction}

At a conference at the Santa Fe Institute in the spring of 1989, John Archibald Wheeler presented an idea he called ``It from Bit,'' in which he proposed that information plays a significant role in the foundation of physics \cite{Wheeler89}. The universe, he suggested, may be fundamentally an information processing system from which the apparent reality of matter somehow emerges. Wheeler might have been led to his conjecture by a 1982 talk by his former graduate student, Richard Feynman, who also pondered the connection between information and quantum physics \cite{Feynman82}. In particular, Feynman discussed whether physics could be simulated by a quantum computer, and suggested that all field theories ``can be simulated in every way, apparently, with little latticeworks of spins and other things.'' He suggested that the appropriate quantum computer might be constructed from what we would now call a quantum cellular automaton (QCA):  ``every finite quantum mechanical system can be described exactly, imitated exactly, by supposing that we have another system such that at each point in space-time this system has only two possible base states. Either that point is occupied or unoccupied...''

Recent work on the connection between quantum walks, quantum cellular automata, and quantum field theory suggests that Feynman and Wheeler may have been right. The qubit is the fundamental unit of quantum information, and quantum information processing is essentially the action of a string of unitary quantum gates on some initial state of qubits. In quantum field theory the time development of a quantum field is given by the action of a unitary operator on a state describing quantum particles, or the creation and annihilation operators that correspond to them. Recent work suggests that systems of the former type, in the continuum limit, yield familiar systems of the latter type that we know from quantum field theory. If this is so, it not only provides insight into the issues raised by Feynman and Wheeler, it offers a new way of understanding quantum field theories. For example, such a QCA could represent an equivalent theory in discrete spacetime that avoids the infinities that plague quantum field theory. Whether discrete spacetime is to be taken literally would then be an interesting question, and may be answerable through technology of the near future, even for lattice spacings on the Planck scale \cite{BrunMlodinow19}.

The first step in elucidating the qubit/QFT connection has been to show that the Dirac equation, which describes the one-particle sector of a field theory, arises as the continuum limit of a quantum walk. Quantum walks  \cite{AharonovY93,Ambainis01,AharonovD01,Kempe03} are unitary analogues of classical random walks.  They can be thought of as discrete models of single-particle dynamics, in which a particle can be located at any vertex of a graph, and at each time step can move along an edge to a neighboring vertex.  This time evolution is a unitary transformation.  Generically, for this unitary transformation to allow nontrivial dynamics, the particle must have an internal degree of freedom (or ``coin'' space) as well as its position degree of freedom \cite{Meyer96a}.  Quantum walks have been widely studied both for their intrinsic interest \cite{Moore02,Yin08,Brun03a,Brun03b,Kendon03,Brun03c,Kendon06} and for their use in algorithms for quantum computers \cite{Childs02,Shenvi03,Ambainis03,Ambainis07,Farhi08,Ambainis10,Reichardt12}.

It has been shown by a variety of researchers using a variety of methods that quantum walks with particular properties on particular lattices can give rise to relativistic wave equations (like the Weyl and Dirac equations) in the long-wavelength limit \cite{Bialynicki94,Meyer96,Strauch06,Bracken07,Chandrashekar10,Chandrashekar11,DAriano14,Arrighi14,Chandrashekar13,Farrelly14,DAriano15,Succi15,Bisio15,Arrighi15,Bisio16,Arrighi16,DAriano17,Raynal17,DAriano17b,MlodinowBrun18,Bisio18,DAriano19,Maeda20,Manighalam19}.  In one recent paper \cite{MlodinowBrun18}, for example, a 3D quantum walk was defined as a product of three coined one-dimensional quantum walks. It was shown, under some simple assumptions such as locality and the absence of a preferred lattice axis or direction, that the time development of the walk leads to the necessity for a four-dimensional internal space (which in the usual Dirac equation implies the existence of anti-matter), a natural maximum to the propagation speed of particles, and a continuum limit that is Lorentz invariant and corresponds to the 3D Dirac equation. That is typical of such analyses:  the long-wavelength limits obey the usual Lorentz symmetry, but this symmetry would be broken at short length scales or high energy  \cite{BrunMlodinow19,Arrighi14b,Bisio17}.

These results are highly suggestive, but leave open an important question.  Relativistic single-particle wave equations are only natural in a very limited sphere of application.  The more natural joining of quantum mechanics and relativity is in quantum field theory \cite{BjorkenDrell65}.  Is it possible to introduce a many-particle quantum theory on discrete spacetime, which recovers the quantum walk on the lattice in the single-particle sector and reproduces the quantum field theory of free fermions in the long-wavelength limit?

As Feynman suggested, cellular automata are the obvious type of system to consider as the generalization of the quantum walk to the many particle sector \cite{Feynman82}.  Just as quantum walks are unitary analogues of random walks, QCAs are unitary analogues of classical cellular automata  \cite{Watrous91,Lloyd93,Watrous95}.  A QCA is a quantum system comprising a regular array of identical subsystems.  These subsystems evolve in discrete time steps by a unitary transformation.  The most subtle requirement---which makes these systems analogous to classical cellular automata---is that this evolution must be local, in the sense that the state of a local subsystem after an evolution step must depend only on its state before the step, and the states of its nearest neighbors.  Satisfying both unitarity and locality requires some care in how an QCA is defined \cite{Meyer96,Grossing88,Fussy93,Durr96,Meyer96b}.  Based on the results deriving relativistic wave equations from quantum walks, there has been significant interest in finding QCAs that give rise to quantum field theories for multiple particles \cite{Bisio15,Raynal17,Bisio15b,Bisio15c,Bisio16b,Mallick16}.

Constructing a QCA that recovers the quantum walk in the one-particle limit can be done in two natural ways: by constructing a quantum lattice gas, or by second quantization \cite{Shakeel13,Farrelly19}. In the quantum lattice gas formulation, particle positions form a lattice of points; a lattice point is a subsystem with internal states representing the presence or absence of particles associated with a lattice direction. Each time step has two parts. In the first, particles jump from their initial lattice point to a neighbor in the associated direction. Next, they interact with the other particles at that new point, leading to new assignments of direction. Models could allow multiple particles of each type to be present, or only on particle; the latter case will generally correspond to ``hard core'' bosons, but could be used to model fermions as well. The quantum lattice gas approach to the Dirac equation is taken in Refs.~\cite{Yepez13,Yepez16}. Here, however, we shall employ the second quantization approach.

In this paper we ``promote'' the 1D quantum walk to derive a 1D QCA that is manifestly both unitary and local. Importantly, our analysis reveals complications that stand in the way of employing a similar straightforward technique to obtain an analogous correspondence in higher dimensions, which is the ultimate goal. In fact, we prove that without some new ingredient, a construction such as we have undertaken is not possible in higher dimension. The issue is another element that Feynman foresaw in his talk.  He said that although he believed that bosons could be described as he proposed, he wasn't sure about fermions.

Fermions present a challenge because they inherently involve a certain kind of nonlocality: creation and annihilation operators anti-commute regardless of their spatial separation. In quantum field theories this does not lead to any physical nonlocality because observables are always sums of products of pairs of creation and annihilation operators, which means that the operators corresponding to observables in separated regions of space commute. But a QCA is written in terms of qubits or spins with local couplings, so the issue is how to account for the nonlocality of the creation and annihilation operators when representing this local system as fermions. In one dimension that can be done using the Jordan-Wigner transformation \cite{Jordan28}, but we prove that no such remedy is possible in higher dimensions.

In 1D, the phase acquired by commuting operators past each other can be produced by an actual physical phase shift, representing a short-range interaction.  But in higher dimensions we show that this does not work for any QCA satisfying a set of reasonable requirements unless the interaction can act at unlimited range.  This is the most important conclusion of this paper.  We discuss possible workaround strategies that can evade this restriction, but all of them require starting with quite different assumptions.

As we were completing the work described in this paper, we became aware of a very interesting recent paper by Arrighi, B\'eny and Farrelly \cite{Arrighi20} that constructs a 1D QCA that also yields a quantum field theory of free fermions in the long-wavelength limit.  That paper also incorporates a local ``gauge'' field to produce a theory with manifest gauge symmetry and considers extensions to the interacting theory.  To include interactions is a key goal of this research program; however, based on the no-go result in our current paper, we believe it will be difficult to generalize these constructions to higher dimensions without a significant change in approach, as we discuss in our conclusions.  The same restriction will apply to lattice gas models like those in \cite{Yepez13,Yepez16}.

\section{The quantum walk in 1D}

\subsection{Quantum walks}

The cellular automaton model that we study in this paper reproduces the behavior of a quantum walk in the one-particle excitation sector.  The quantum walk model in one dimension is given by a particle on a one-dimensional lattice of points $\ket{x}$ spaced a distance $\Delta x$ apart from each other \cite{MlodinowBrun18}.  The particle also has an internal degree of freedom, or ``coin space.''  The Hilbert space has the form $\mathcal{H} = \mathcal{H}_X \otimes \mathcal{H}_C$.  In the quantum walk, time is discrete, with the timesteps separated by $\Delta t$, and the evolution from one time to the next is given by the time-evolution unitary:
\begin{equation}
\label{eq:QWalkEvolution}
\ket{\psi_{t+\Delta t}} = U\ket{\psi_t} = \left( I\otimes C \right) \left( \sum_{j=\pm} S_j \otimes P_j \right) \ket{\psi_t} ,
\end{equation}
where the $\{S_j\}$ are shift operators that move the particle from its current position to its neighbor in the direction $j$.  The $\{P_j\}$ are orthogonal projectors on the internal space; and $C$ is a unitary that acts on the internal space, often called the ``coin flip'' unitary.

The idea is that the walk proceeds by a process analogous to a series of coin flips.  The projectors $\{P_\pm\}$ correspond to different faces of the coin, which indicate which direction to move (plus and minus, or right and left); the unitary $C$ scrambles the faces, so that one does not constantly move in the same direction.  But in the unitary case, unlike classical random walks, the evolution is always invertible, and interference effects are very important.  For the present we will assume that the internal space is two-dimensional, spanned by a basis $\{\ket{R}, \ket{L}\}$, indicating the directions ``right'' and ``left,'' respectively; this could be generalized if necessary, but is sufficient in 1D.

For the quantum walk on a 1D lattice \cite{MlodinowBrun18}, the time evolution operator takes the form
\begin{equation}
\label{eq:OneDWalk}
U = \left( I\otimes C \right) \left( S \otimes \ket{R}\bra{R} + S^\dagger \otimes \ket{L}\bra{L} \right) ,
\end{equation}
where the shift operator acts as $S \ket{x} = \ket{x+\Delta x}$ and $S^\dagger \ket{x} = \ket{x-\Delta x}$,  and $C$ is a $2\times2$ unitary matrix.   We will generally consider families of coin-flip unitaries, parametrized as rotations in Hilbert space:
\begin{equation}
C(\theta) = e^{-i \theta Q } ,
\end{equation}
where $Q$ is a Hermitian operator that acts on the internal space.  $Q$ flips the $R$ and $L$ directions.  We can, without loss of generality, assume that $Q^2 = I$ and $\tr\{Q\} = 0$.

\subsection{The momentum representation and continuous limit}

As shown in \cite{MlodinowBrun18} we can transform the position degree of freedom to a momentum representation.  The eigenvectors of the shift operators take the form
\begin{equation}
\label{eq:1DMomentumStates}
\ket{k} = \sum_{j=-\infty}^\infty e^{-i k j\Delta x} \ket{j\Delta x} ,\ \ \ \ -\pi < k\Delta x \le \pi ,
\end{equation}
which have eigenvalues
\begin{equation}
S\ket{k} = e^{i k\Delta x} \ket{k} ,\ \ \ \ S^\dagger\ket{k} = e^{-i k\Delta x} \ket{k} .
\end{equation}
These momentum states $\{\ket{k}\}$ are not normalizable, but it is possible to write a normalizable wavefunction in terms of these eigenstates using the inverse transform
\begin{equation}
\ket{x} = \frac{1}{2\pi} \int_{-\pi}^{\pi} dk\, e^{i k x} \ket{k} ,
\end{equation}
where $x = j\Delta x$ for some integer $-\infty < j < \infty$.

Writing the evolution operator (\ref{eq:OneDWalk}) in terms of momentum yields a compact form:
\begin{eqnarray}
\label{eq:OneDWalkMomentum}
U &=& e^{-i\theta Q} \left( e^{i k\Delta x} P_R + e^{-i k\Delta x} P_L \right) \nonumber\\
&=& e^{-i\theta Q} \left( \cos(k\Delta x) (P_R + P_L) +i\sin(k\Delta x)(P_R - P_L) \right) \nonumber\\
&=& e^{-i\theta Q} \left( \cos(k\Delta x) I +i\sin(k\Delta x) \Delta P \right) \nonumber\\
&=& e^{-i \theta Q} e^{k\Delta x \Delta P} ,
\end{eqnarray}
where $\Delta P = P_R - P_L = \ket{R}\bra{R} - \ket{L}\bra{L}$.  We can straightforwardly go to the continuum (long-wavelength) limit $|k\Delta x| \ll 1$ and $\theta \ll 1$. 
Suppose that the time between steps is $\Delta t$.  Then we can expand the two exponentials and retain only terms linear in $k\Delta x$ and $\theta$:
\begin{eqnarray}
\label{eq:Cont1Dmomentum}
\partial_t \ket\psi &\equiv& \frac{\ket{\psi(t+\Delta t)} - \ket{\psi(t)}}{\Delta t} = i\left( ( k\Delta x/\Delta t) \Delta P - (\theta/\Delta t) Q \right) \ket\psi \nonumber\\
&=& \frac{i}{\hbar}\left( pc \Delta P - mc^2 Q \right) \ket\psi ,
\end{eqnarray}
where we have defined $p \equiv \hbar k$, $c \equiv \Delta x/\Delta t$ and $m \equiv \hbar\theta/c^2 \Delta t$.  This is exactly the Dirac equation for one spatial dimension in momentum form, where $\Delta P$ and $Q$ play the role of $2\times2$ gamma matrices (i.e., Pauli matrices).

\section{The 1D cellular automaton model}

\subsection{The many-particle Hilbert space}

How do we generalize the quantum walk to multiple particles?  The natural way to do this is with a {\it quantum cellular automaton} model.  Now, instead of describing a single particle with a position degree of freedom and an internal degree of freedom, we have many local subsystems, arranged in a 1D lattice with spacing $\Delta x$:
\begin{equation}
\mathcal{H} = \cdots \otimes \mathcal{H}_{x-\Delta x} \otimes \mathcal{H}_{x} \otimes \mathcal{H}_{x+\Delta x} \otimes \cdots
\end{equation}
(This Hilbert space is nonseparable, but we consider only the subspace with a finite number of particles, which is a separable space.)  Each subsystem in the above equation (say at location $x$) has a four-dimensional Hilbert space $\mathcal{H}_x$ with basis states $\{ \ket{j_-}_{x-} \otimes \ket{j_+}_{x+}$, where $j_\pm = 0, 1$.  These states can be interpreted as 0 or 1 particles at location $x$ with internal state $+$ or $-$.  This means that the local Hilbert space further factorizes into $\mathcal{H}_x = \mathcal{H}_{x,-} \otimes \mathcal{H}_{x,+}$, where $\mathcal{H}_{x,\pm} = \mathbb{C}^2$ is the Hilbert space of a single qubit.

For brevity, we can represent the local basis states of $\mathcal{H}_x$ by 2-bit strings:  $\{ \ket{00}, \ket{01}, \ket{10}, \ket{11} \}$.  The basis vectors of the entire lattice, then, can be written as long binary strings:
\begin{equation}
\ket{ \cdots j_{x}^{-}\, j_{x} ^{+}\, j_{x+\Delta x}^{-}\, j_{x+\Delta x}^{+} \cdots} .
\end{equation}
If the total number of particles is finite---that is, the total number of sites where $j_x^\pm = 1$ is finite---then we can use a short-hand notation for basis states:
\begin{equation}
\label{eq:orderingRep}
\ket{ x_{1},\pm; x_{2},\pm; \ldots; x_{n},\pm} ,
\end{equation}
where the values $x_1,\ldots,x_n$ are the locations of the $n$ particles, and $\pm$ indicates the internal state.  Our convention is that we list the locations in ascending order from left to right, so $x_1 \le x_2 \le \cdots \le x_n$.  If there are two particles at the same site (i.e., with the same value of $x$), our convention is that we list $x,-$ first and $x,+$ second.  We write the vacuum (no particle) state as $\ket\Omega \equiv \ket{\cdots 000 \cdots}$.

In this paper we will consider the Fock space comprising all states of $n$ particles for all $n = 0, 1, 2, \ldots$.  The Fock space decomposes into subspaces of different particle number:
\[
\mathcal{H}^{\rm Fock} = \mathcal{H}^0 \oplus \mathcal{H}^1 \oplus \mathcal{H}^2 \oplus \cdots
  = \bigoplus_{n=0}^\infty \mathcal{H}^n ,
\]
where $\mathcal{H}^n$ is the subspace of all $n$-particle states.

\subsection{Time evolution}

How do states evolve in this model?  Just as in the quantum walk, time is discrete, and the unitary time evolution operator is the product of two operators:  an operator $\hat{\Sigma}$ that shifts all the $+$ particles in the $+$ direction and the $-$ particles in the $-$ direction, and an operator $\hat{C}$ that rotates the internal state:
\begin{equation}
\hat{U} = \hat{C} \hat{\Sigma} .
\end{equation}
We will define $\hat{C}$ and $\hat\Sigma$ one at a time below.  It is important to emphasize that this evolution operator $U$ acts {\it locally}:  that is, any observables acting on a site $x$ after applying $U$ can only depend on observables at neighboring sites before applying $U$.  This locality is a requirement for a cellular automaton model.  We need to remember this, because in deriving quantum field theory as a continuum limit, we will express the theory in terms of creation and annihilation operators that are {\it not} local in this sense.  In spite of this representation in terms of nonlocal operators, the underlying theory will be local.

To make this locality very clear, we will define the two operators $\hat{\Sigma}$ and $\hat{C}$ to each be the tensor product of operators acting only on local subsystems.  As can be seen in the definitions that follow, each on its own would give only trivial evolution, in which particles cannot propagate; but by alternating them, nontrivial propagation becomes possible.  Also, note that the vacuum state is invariant: $\hat{U}\ket\Omega = \ket\Omega$.

\subsubsection{The coin operator}

The coin operator is a tensor product of unitary operators acting on the local four-dimensional spaces $\mathcal{H}_x$:
\begin{equation}
\hat{C} = \cdots \otimes C_{x-\Delta x} \otimes C_x \otimes C_{x+\Delta x} \otimes \cdots ,
\end{equation}
where each of these local unitaries $C_x$ acts on the basis vectors as follows:
\begin{eqnarray}
\label{eq:coinOpDef}
C \ket{00} &=& \ket{00}, \nonumber\\
C \ket{01} &=& \cos(\theta)\ket{10} + \sin(\theta)\ket{01}, \\
C \ket{10} &=& \cos(\theta) \ket{01} - \sin(\theta)\ket{10}, \nonumber\\
C \ket{11} &=& -\ket{11}, \nonumber
\end{eqnarray}
where $\theta$ is once again a dimensionless parameter that will be assumed to be small.  At $\theta=0$, the unitary $C$ just switches the basis vectors $\ket{01} \leftrightarrow \ket{10}$ (corresponding to a single particle at $x$ in the $+$ or $-$ internal state) and applies a phase of $-1$ to $\ket{11}$.

\subsubsection{The shift operator}

The shift operator can also be written as a tensor product of unitaries acting on local subystems,
\begin{equation}
\hat{\Sigma} = \cdots \otimes S_{x-\Delta x} \otimes S_x \otimes S_{x+\Delta x} \otimes \cdots
\end{equation}
But rather that acting on the local Hilbert space $\mathcal{H}_x = \mathcal{H}_{x,-} \otimes \mathcal{H}_{x,+}$, these unitaries $S_x$ are offset, and act on the space $\mathcal{H}_{x,+} \otimes \mathcal{H}_{x+\Delta x,-}$.  On the basis vectors of this space, the local unitary acts as follows:
\begin{eqnarray}
\label{eq:shiftOpDef}
S\ket{00} &=& \ket{00}, \nonumber\\
S\ket{01} &=& \ket{10}, \\
S\ket{10} &=& \ket{01}, \nonumber\\
S\ket{11} &=& -\ket{11}. \nonumber
\end{eqnarray}
This is essentially the same action as the local coin-flip unitary $C$ for $\theta=0$, but offset.  In that limit, the evolution $\hat{U}$ of the cellular automaton is very simple:  particles in the internal state $\ket+$ move to the right by $\Delta x$ at each time step, while particles in the internal state $\ket-$ move to the left.

This way of defining the time evolution $\hat{U}$ has three interesting properties:
\begin{enumerate}
\item The evolution operator $\hat{U}$ is both unitary (since it is the product of two unitaries $\hat{C}$ and $\hat{\Sigma}$) and strictly local, as required to be a cellular automaton.

\item The single-particle sector of this cellular automaton evolves like the 1D quantum walk defined above for $\Delta P = \sigma_3$ and $Q = \sigma_2$ (as we will show below).

\item The minus sign in the action of $C$ and $S$ on the $\ket{11}$ basis states means that when two particles pass each other, the state acquires a phase of $-1$.  This will allow us to define creation and annihilation operators for this model that obey Fermi statistics (as we will also show below).
\end{enumerate}

\subsection{The one-particle sector and quantum walks}

The basis states corresponding to a single particle are $\ket{x,\pm}\in \mathcal{H}^1$.  Formally we can write this Hilbert space as a tensor product of a position space and an internal (coin) space:
\begin{equation}
\mathcal{H}^1 = \mathcal{H}_{\rm pos} \otimes \mathcal{H}_{\rm coin}
= \mathbb{C}^\infty \otimes \mathbb{C}^2 .
\end{equation}
The basis vectors of $\mathcal{H}_{\rm pos}$ are $\{\ket{x} \equiv \ket{j\Delta x}\}$ and the basis vectors of $\mathcal{H}_{\rm coin}$ are $\ket{\pm}$, which we can identify with $\ket{R}$ and $\ket{L}$ from Eq.~(\ref{eq:OneDWalk}).

Since both the shift operator $\hat{\Sigma}$ and the coin operator $\hat{C}$ conserve particle number, the evolving state will remain in $\mathcal{H}^1$ at all times.  How does this basis state evolve under $\hat{U}$?
\begin{equation}
\hat{U} \ket{x,\pm} = \cos(\theta) \ket{x \pm \Delta x,\pm}  \pm \sin(\theta) \ket{x \pm \Delta x,\mp} .
\end{equation}
It is quite straightforward to see that this is precisely how the quantum walk given by Eq.~(\ref{eq:OneDWalk}) evolves if we choose $Q = i\left( \ket{L}\bra{R} - \ket{R}\bra{L} \right) \equiv \sigma_2$.  Thus we see that the one-particle sector of this quantum cellular automaton model reproduces the results of the quantum walk on the 1D lattice.

\subsection{Creation and annihilation operators}

We will now define a set of creation and annihilation operators $\{ \hat{a}^\dagger_{x,\varepsilon}\}$ and $\{ \hat{a}_{x,\varepsilon}\}$ where $x = j\Delta x$ for some integer $j$ and $\varepsilon = \pm$, which create (or destroy) particles at a particular location $x$ in the internal state $\varepsilon$.  These operators will be defined to obey the usual anticommutation relations, and to evolve under the time evolution unitary in a simple way.

\subsubsection{Definition and ordering}

First, in acting on the vacuum state the creation operator gives
\begin{equation}
\label{eq:creationVacuum}
\hat{a}^\dagger_{x,\varepsilon}\ket\Omega = \ket{x,\varepsilon} ,
\end{equation}
and any annihilation operator gives $\hat{a}_{x,\varepsilon}\ket\Omega = 0$.

Acting with $\hat{a}^\dagger_{x,\varepsilon}$ on an $n$-particle basis state $\ket{x_1,\varepsilon_1; \cdots; x_n, \varepsilon_n}$ gives
\begin{equation}
\label{eq:creationDef}
\hat{a}^\dagger_{x,\varepsilon} \ket{x_1,\varepsilon_1; \cdots; x_n, \varepsilon_n} =
{(-1)}^m \ket{x_1,\varepsilon_1; \cdots; x_m, \varepsilon_m; x, \varepsilon; \cdots; x_n, \varepsilon_n},
\end{equation}
if the site $x,\varepsilon$ is not already occupied in the original state; if it is, then
\[
\hat{a}^\dagger_{x,\varepsilon} \ket{x_1,\varepsilon_1; \cdots; x_n, \varepsilon_n} = 0 .
\]
The value $m$ in Eq.~(\ref{eq:creationDef}) is determined by the ordering convention from Eq.~(\ref{eq:orderingRep}).  With this definition, we can readily see that
\begin{equation}
\label{eq:createBasisVec}
\ket{x_1,\varepsilon_1; \cdots; x_n, \varepsilon_n} = \hat{a}^\dagger_{x_1,\varepsilon_1} \cdots \hat{a}^\dagger_{x_n,\varepsilon_n} \ket\Omega .
\end{equation}
The definition of the annihilation operator $\hat{a}_{x,\varepsilon}$, which is the Hermitian conjugate of the creation operator, follows from Eq.~(\ref{eq:creationVacuum}) and Eq.~(\ref{eq:creationDef}).  Since these operators anticommute even when they create (or destroy) particles that are far apart, they act nontrivially throughout space.  These creation and annihilation operators therefore are highly nonlocal, even though they create (or destroy) local excitations.  This does not contradict the locality of the theory, since the underlying QCA is still local.

\subsubsection{Anticommutation relations}

From the definition above, we can see that on any basis state $\ket{b}$
\begin{equation}
\hat{a}^\dagger_{x,\varepsilon} \hat{a}^\dagger_{x',\varepsilon'} \ket{b} = - \hat{a}^\dagger_{x',\varepsilon'} \hat{a}^\dagger_{x,\varepsilon} \ket{b} ,
\end{equation}
which implies that $\hat{a}^\dagger_{x,\varepsilon}$ and $\hat{a}^\dagger_{x',\varepsilon'}$ anticommute:
\[
\{\hat{a}^\dagger_{x,\varepsilon},\hat{a}^\dagger_{x',\varepsilon'}\} = 0 .
\]
This obviously immediately implies that $\{\hat{a}_{x,\varepsilon},\hat{a}_{x',\varepsilon'}\} = 0$ as well.

The mixed anticommutation relation is slightly less obvious, but the standard relation still holds.  For $(x,\varepsilon) \ne (x',\varepsilon')$, $\hat{a}^\dagger_{x,\varepsilon}$ and $\hat{a}_{x',\varepsilon'}$ anticommute.  When they are equal, we can see that for any basis state $\ket{b}$, either
\[
\hat{a}^\dagger_{x,\varepsilon} \hat{a}_{x,\varepsilon} \ket{b} = \ket{b} \ \ {\rm and}\ \ 
\hat{a}_{x,\varepsilon} \hat{a}^\dagger_{x,\varepsilon} \ket{b} = 0 ,
\]
if the site $x,\varepsilon$ is occupied in $\ket{b}$, or
\[
\hat{a}^\dagger_{x,\varepsilon} \hat{a}_{x,\varepsilon} \ket{b} = 0 \ \ {\rm and}\ \ 
\hat{a}_{x,\varepsilon} \hat{a}^\dagger_{x,\varepsilon} \ket{b} = \ket{b} ,
\]
if it is not.  Putting these all together gives us the final anticommutation relation:
\[
\{\hat{a}^\dagger_{x,\varepsilon},\hat{a}_{x',\varepsilon'}\} = 
\delta_{xx'} \delta_{\varepsilon \varepsilon'} I .
\]

\subsubsection{Time evolution}

Based on the relationship in Eq.~(\ref{eq:createBasisVec}) between basis vectors and creation operators, we can see how these creation and annihilation operators evolve under the time-evolution operator $\hat{U}$:
\begin{eqnarray}
\label{eq:basisStateEvolution}
\hat{U} \ket{x_1,\varepsilon_1; \cdots; x_n, \varepsilon_n} &=& \hat{U} \hat{a}^\dagger_{x_1,\varepsilon_1} \cdots \hat{a}^\dagger_{x_n,\varepsilon_n} \ket\Omega \nonumber\\
&=& \hat{U} \hat{a}^\dagger_{x_1,\varepsilon_1} \hat{U}^\dagger \hat{U} \cdots \hat{U}^\dagger \hat{U} \hat{a}^\dagger_{x_n,\varepsilon_n} \hat{U}^\dagger \hat{U} \ket\Omega \nonumber\\
&=& \left( \hat{U} \hat{a}^\dagger_{x_1,\varepsilon_1} \hat{U}^\dagger \right) \cdots \left( \hat{U} \hat{a}^\dagger_{x_n,\varepsilon_n} \hat{U}^\dagger \right) \ket\Omega ,
\end{eqnarray}
where the last step uses the invariance of the vacuum.  Referring back to Eq.~(\ref{eq:coinOpDef}) and Eq.~(\ref{eq:shiftOpDef}), we see that every time a pair of particles passes each other the state acquires a phase of $-1$; this matches the phase of $-1$ from reordering two creation operators.

Here are some simple examples for $n=2$.  First, consider two particles on neighboring sites moving in opposite directions:
\begin{eqnarray}
\label{eq:n2example1}
\hat{U} \ket{x,+; x+\Delta x,-} &=& -\cos^2(\theta) \ket{x,-; x+\Delta x,+} - \cos(\theta)\sin(\theta) \ket{x,-; x+\Delta x,-} \nonumber\\
&& + \cos(\theta)\sin(\theta) \ket{x,+; x+\Delta x,+} + \sin^2(\theta) \ket{x,+; x+\Delta x,-} \nonumber\\
&=& -\cos^2(\theta) \hat{a}^\dagger_{x,-} \hat{a}^\dagger_{x+\Delta x,+}\ket\Omega - \cos(\theta)\sin(\theta) \hat{a}^\dagger_{x,-} \hat{a}^\dagger_{x+\Delta x,-}\ket\Omega \\
&& + \cos(\theta)\sin(\theta) \hat{a}^\dagger_{x,+} \hat{a}^\dagger_{x+\Delta x,+}\ket\Omega + \sin^2(\theta) \hat{a}^\dagger_{x,+} \hat{a}^\dagger_{x+\Delta x,-}\ket\Omega \nonumber\\
&=& \cos^2(\theta) \hat{a}^\dagger_{x+\Delta x,+} \hat{a}^\dagger_{x,-} \ket\Omega + \cos(\theta)\sin(\theta)  \hat{a}^\dagger_{x+\Delta x,-} \hat{a}^\dagger_{x,-} \ket\Omega \nonumber\\
&& - \cos(\theta)\sin(\theta) \hat{a}^\dagger_{x+\Delta x,+} \hat{a}^\dagger_{x,+} \ket\Omega - \sin^2(\theta) \hat{a}^\dagger_{x+\Delta x,-} \hat{a}^\dagger_{x,+} \ket\Omega \nonumber\\
&=& \left( \cos(\theta) \hat{a}^\dagger_{x+\Delta x,+} + \sin(\theta) \hat{a}^\dagger_{x+\Delta x,-} \right)
\left( \cos(\theta) \hat{a}^\dagger_{x,-} - \sin(\theta) \hat{a}^\dagger_{x,+} \right) \ket\Omega , \nonumber
\end{eqnarray}
where in the third equality we used the anticommutation of creation operators.  Now consider two particles moving onto the same site from opposite directions:
\begin{eqnarray}
\label{eq:n2example2}
\hat{U} \ket{x-\Delta x,+; x+\Delta x,-} &=& - \ket{x,-; x,+}  
= - \hat{a}^\dagger_{x,-} \hat{a}^\dagger_{x,+}\ket\Omega \\
&=& - \left( \cos^2(\theta) + \sin^2(\theta) \right) \hat{a}^\dagger_{x,-} \hat{a}^\dagger_{x,+} \ket\Omega \nonumber\\
&=& \left( \cos^2(\theta) \hat{a}^\dagger_{x,+} \hat{a}^\dagger_{x,-} - \sin^2(\theta) \hat{a}^\dagger_{x,-} \hat{a}^\dagger_{x,+} \right) \ket\Omega \nonumber\\
&=& \left( \cos(\theta) \hat{a}^\dagger_{x,+} + \sin(\theta) \hat{a}^\dagger_{x,-} \right)
\left( \cos(\theta) \hat{a}^\dagger_{x,-} - \sin(\theta) \hat{a}^\dagger_{x,+} \right) \ket\Omega ,\nonumber
\end{eqnarray}
where in the fourth equality we used anticommutation and in the last we added and subtracted terms (or alternatively, the fact that ${\left(\hat{a}^\dagger_{x,\varepsilon}\right)}^2 = 0$).

Generalizing from Eqs.~(\ref{eq:basisStateEvolution}), (\ref{eq:n2example1}) and (\ref{eq:n2example2}), and taking Hermitian conjugates to get the results for annihilation operators, we see that
\begin{eqnarray}
\label{eq:timeEvolvePos}
\hat{U} \hat{a}^\dagger_{x,+} \hat{U}^\dagger &=& \cos(\theta) \hat{a}^\dagger_{x+\Delta x,+} + \sin(\theta) \hat{a}^\dagger_{x+\Delta x,-} , \nonumber\\
\hat{U} \hat{a}^\dagger_{x,-} \hat{U}^\dagger &=& \cos(\theta) \hat{a}^\dagger_{x-\Delta x,-} - \sin(\theta) \hat{a}^\dagger_{x-\Delta x,+} , \nonumber\\
\hat{U} \hat{a}_{x,+} \hat{U}^\dagger &=& \cos(\theta) \hat{a}_{x+\Delta x,+} + \sin(\theta) \hat{a}_{x+\Delta x,-} , \nonumber\\
\hat{U} \hat{a}_{x,-} \hat{U}^\dagger &=& \cos(\theta) \hat{a}_{x-\Delta x,-} - \sin(\theta) \hat{a}_{x-\Delta x,+} .
\label{eq:positionEvolve}
\end{eqnarray}
So creation and annihilation operators evolve into simple linear combinations of themselves under the time evolution $\hat{U}$. This is only true because of the dynamical phases introduced in Eq.~(\ref{eq:coinOpDef}) and Eq.~(\ref{eq:shiftOpDef}) that flip the sign of the state when two particles pass each other; without those phases, creation and annihilation operators would not evolve into linear combinations of themselves as in Eq.~(\ref{eq:positionEvolve}).

\section{The momentum and energy representations and the long-wavelength limit}

\subsection{Momentum picture}

Having defined creation and annihilation operators in position as above, we can define creation and annihilation operators in the {\it momentum} picture straightforwardly:
\begin{equation}
\label{eq:opMomDef}
a_{\pm k,\pm} = \sum_{j=-\infty}^\infty e^{ijk\Delta x} \hat{a}_{j\Delta x,\pm} ,\ \ \ \ 
a^\dagger_{\pm k,\pm} = \sum_{j=-\infty}^\infty e^{-ijk\Delta x} \hat{a}^\dagger_{j\Delta x,\pm} ,\ \ \ \ 
-\frac{\pi}{\Delta x} < k \le \frac{\pi}{\Delta x} .
\end{equation}
From the anticommutation relations in the position representation, we can easily show that
\[
\{ a_{k,\varepsilon}, a_{k',\varepsilon'} \} = \{ a^\dagger_{k,\varepsilon}, a^\dagger_{k',\varepsilon'} \} = 0 ,\ \ \ \ 
\{ a^\dagger_{k,\varepsilon}, a_{k',\varepsilon'} \} = \delta_{\varepsilon,\varepsilon'} \delta(k-k') I ,
\]
where $\varepsilon,\varepsilon' = \pm$.  In particular, this implies that $( a_{k,\varepsilon})^2 = ( a^\dagger_{k,\varepsilon})^2 = 0$.  We can interpret this as implying that no more than one particle can occupy a momentum state $k,\varepsilon$.

Drawing on the results from Eq.~(\ref{eq:timeEvolvePos}), we can see how these operators transform under the cellular automaton evolution:
\begin{eqnarray}
\label{eq:timeEvolveMom1}
\hat{U} a_{\pm k,\pm} \hat{U}^\dagger &=& \sum_{j=-\infty}^\infty e^{ijk\Delta x} \hat{U}\hat{a}_{j\Delta x,\pm} \hat{U}^\dagger \\
&=& \sum_{j=-\infty}^\infty e^{ijk\Delta x} \left( \cos(\theta) \hat{a}_{(j\pm1)\Delta x,\pm} \pm \sin(\theta) \hat{a}_{(j\pm1)\Delta x,\mp} \right) \nonumber\\
&=& e^{\mp ik\Delta x} \sum_{j=-\infty}^\infty e^{ijk\Delta x} \left( \cos(\theta) \hat{a}_{j\Delta x,\pm} \pm \sin(\theta) \hat{a}_{j\Delta x,\mp} \right) \nonumber\\
&=& e^{\mp ik\Delta x} \left( \cos(\theta) a_{\pm k,\pm} \pm \sin(\theta) a_{\mp k,\mp} \right) , \nonumber
\end{eqnarray}
\begin{eqnarray}
\label{eq:timeEvolveMom2}
\hat{U} a^\dagger_{\pm k,\pm} \hat{U}^\dagger &=& \sum_{j=-\infty}^\infty e^{-ijk\Delta x} \hat{U}\hat{a}^\dagger_{j\Delta x,\pm}\hat{U}^\dagger \\
&=& \sum_{j=-\infty}^\infty e^{-ijk\Delta x} \left( \cos(\theta) \hat{a}^\dagger_{(j\pm1)\Delta x,\pm} \pm \sin(\theta) \hat{a}^\dagger_{(j\pm1)\Delta x,\mp} \right) \nonumber\\
&=& e^{\pm ik\Delta x} \sum_{j=-\infty}^\infty e^{-ijk\Delta x} \left( \cos(\theta) \hat{a}^\dagger_{j\Delta x,\pm} \pm \sin(\theta) \hat{a}^\dagger_{j\Delta x,\mp} \right) \nonumber\\
&=& e^{\pm ik\Delta x} \left( \cos(\theta) a^\dagger_{\pm k,\pm} \pm \sin(\theta) a^\dagger_{\mp k,\mp} \right) . \nonumber
\end{eqnarray}

\subsection{Energy picture}

In Eq.~(\ref{eq:timeEvolveMom1}) and (\ref{eq:timeEvolveMom2}) we observe an interesting difference between the momentum and position representations.  In position, the time evolution shifts creation and annihilation operators to the left or right, so the ordering of a product of operators can change.  By contrast, in momentum, the time evolution mixes the $k,+$ and $-k,-$ states, but otherwise leaves the momentum unchanged.  In other words, $\hat{U}$ acts independently on particles of different momenta. Let's look at how linear combinations of these operators transform:
\begin{eqnarray}
\hat{U} \left( \alpha a_{k,+} + \beta a_{-k,-} \right) \hat{U}^\dagger
  &=& \left( e^{-ik\Delta x} \cos(\theta) \alpha - e^{ik\Delta x} \sin(\theta) \beta \right) a_{k,+} \nonumber\\
&& +  \left( e^{ik\Delta x} \cos(\theta) \beta + e^{-ik\Delta x} \sin(\theta) \alpha \right) a_{-k,-} ,
\end{eqnarray}
\begin{equation}
\label{eq:momentumRotation}
\Rightarrow \left(\begin{array}{c} \alpha \\ \beta \end{array}\right) \longrightarrow 
  \left(\begin{array}{cc} e^{-ik\Delta x} \cos(\theta) & - e^{ik\Delta x} \sin(\theta) \\
    e^{-ik\Delta x} \sin(\theta) & e^{ik\Delta x} \cos(\theta) \end{array} \right)
    \left(\begin{array}{c} \alpha \\ \beta \end{array}\right) 
\equiv \mathbf{M} \left(\begin{array}{c} \alpha \\ \beta \end{array}\right) .
\end{equation}
By diagonalizing the matrix $\mathbf{M}$ in Eq.~(\ref{eq:momentumRotation}) we can find a new pair of operators that are invariant under $\hat{U}$ up to a phase.  The eigenvalues of $\mathbf{M}$ are
\begin{equation}
\lambda_{k,\pm} \equiv e^{\pm i\phi_k} = \cos(\theta) \cos(k\Delta x) \pm i\sqrt{1- \cos^2(\theta) \cos^2(k\Delta x)} .
\label{eq:phaseDef}
\end{equation}
Let the corresponding eigenvectors be
\begin{equation}
\mathbf{v}_{k,\pm} \equiv \left(\begin{array}{c} \alpha_{k,\pm} \\ \beta_{k,\pm} \end{array} \right)
= \frac{1}{\mathcal{N}_\pm} \left(\begin{array}{c} \sin(k\Delta x) \cos(\theta) \mp
  \sqrt{1- \cos^2(\theta) \cos^2(k\Delta x)} \\
  i e^{-i k\Delta x} \sin(\theta) \end{array} \right) ,
\end{equation}
where $\mathcal{N}_\pm$ is a normalization factor.   We can rewrite $\mathbf{M}$ as
\begin{eqnarray}
\mathbf{M} &=& \cos(\theta) \cos(k\Delta x) I - i\bigl( \sin(\theta)\sin(k\Delta x) \sigma_1 \nonumber\\
&& + \sin(\theta)\cos(k\Delta x) \sigma_2 + \cos(\theta)\sin(k\Delta x) \sigma_3 \bigr) \nonumber\\
&\equiv& \cos(\phi_k) I - i \sin(\phi_k) \hat{n}\cdot\vec{\sigma} ,
\end{eqnarray}
where $\hat{n}$ is a real unit 3-vector, and from Eq.~(\ref{eq:phaseDef})
\begin{equation}
\phi_k = \arccos\left( \cos(\theta) \cos(k\Delta x) \right) .
\end{equation}

We can define a new set of creation and annihilation operators:
\begin{equation}
b_{k,\pm} = \alpha_{k,\pm} a_{k,+} + \beta_{k,\pm} a_{k,-} ,
\end{equation}
which obey the usual anticommutation relations
\[
\{ b_{k,\varepsilon}, b_{k',\varepsilon'} \} = \{ b^\dagger_{k,\varepsilon}, b^\dagger_{k',\varepsilon'} \} = 0 ,\ \ \ \ 
\{ b^\dagger_{k,\varepsilon}, b_{k',\varepsilon'} \} = \delta_{\varepsilon,\varepsilon'} \delta(k-k') I ,
\]
and which evolve in time just by a simple phase rotation:
\begin{equation}
\hat{U} b_{k,\pm} \hat{U}^\dagger = e^{\pm i\phi_k} b_{k,\pm} .
\end{equation}
This trivial time evolution suggests a way of defining a set of $n$-particle ``energy'' eigenstates:
\begin{equation}
\ket{k_1,\varepsilon_1; k_2,\varepsilon_2; \cdots; k_n,\varepsilon_n} \equiv b^\dagger_{k_1,\varepsilon_1} \cdots b^\dagger_{k_n,\varepsilon_n} \ket\Omega ,
\end{equation}
where $\varepsilon_i = \pm$.  We choose a simple ordering $k_1 \le k_2 \le \cdots \le k_n$.  If both $k,-$ and $k,+$ are present, we list $k,-$ first.  It is quite easy to see that these (unnormalizable) basis states are orthogonal to each other, and that any state can be written as an integral over them.  Moreover, they are eigenstates of the time evolution unitary:
\begin{equation}
\hat{U} \ket{k_1,\varepsilon_1; k_2,\varepsilon_2; \cdots; k_n,\varepsilon_n} =
\exp\left( i \sum_{j=1}^n \varepsilon_j \phi_{k_j} \right) 
\ket{k_1,\varepsilon_1; k_2,\varepsilon_2; \cdots; k_n,\varepsilon_n} .
\end{equation}
It is natural to relate these phases to energies.  If the duration of each time step is $\Delta t$, then we can define the energy of a particle by
\begin{equation}
\pm \phi_k \equiv \mp E_k \Delta t ,
\end{equation}
where we are implicitly taking $\hbar = 1$.  So this system includes both positive and negative energy states.

\subsection{The long-wavelength limit}

In the quantum walk (single particle) case, we recovered the Dirac equation in the long-wavelength limit; that is, when $|k\Delta x| \ll 1$ and also $|\theta| \ll 1$.  With the solutions above, this limit takes a highly suggestive form.  The eigenvalues to first order become
\begin{equation}
\lambda_{k,\pm} \equiv e^{\pm i\phi_k} \approx 1 \pm i \sqrt{\theta^2 + k^2 \Delta x^2} ,
\end{equation}
which gives us an expression for the energy
\begin{equation}
E_k \approx \frac{1}{\Delta t} \sqrt{\theta^2 + k^2 \Delta x^2} \equiv \sqrt{p^2 c^2 + m^2 c^4} ,
\end{equation}
where we have made the same identification as in \cite{MlodinowBrun18} of the speed of light $c = \Delta x/\Delta t$, momentum $p = k$, and particle mass $m = \theta\Delta t/\Delta x^2$.  This is the usual relativistic expression for the energy of a free particle with mass $m$ and momentum $p$.

We can define a kind of Hamiltonian picture of the dynamics using these energy states:
\begin{equation}
\hat{U} = \exp\{ -i \hat{H} \Delta t \} ,
\end{equation}
where the operator $\hat{H}$ is
\begin{equation}
\hat{H} = \frac{1}{\Delta t} \int dk E_k ( b^\dagger_{k,+} b_{k,+} - b^\dagger_{k,-} b_{k,-} ) .
\end{equation} 
We can rewrite this in vector form:
\begin{equation}
\hat{H} =  \int dk \left( \begin{array}{cc} b^\dagger_{k,+} & b^\dagger_{k,-} \end{array}\right)
(E_k/\Delta t) \sigma_3 \left( \begin{array}{c} b_{k,+} \\ b_{k,-} \end{array}\right) .
\end{equation}
This can, in turn, be written in terms of the creation and annihilation operators for momentum.  If we orthogonally diagonalize the $2\times2$ matrix $\mathbf{M}$ in Eq.~(\ref{eq:momentumRotation}),
\begin{equation}
\mathbf{M} = \mathbf{T} \left( \begin{array}{cc} e^{-i\phi_k} & 0 \\ 0 & e^{i\phi_k} \end{array}\right) \mathbf{T}^\dagger,\ \ \ 
\mathbf{T} = \left( \begin{array}{cc} \mathbf{v}_{k,+} & \mathbf{v}_{k,-} \end{array}\right) .
\end{equation}
We can then write the expression for $\hat{H}$ as
\begin{equation}
\hat{H} =  \int dk \left( \begin{array}{cc} a^\dagger_{k,+} & a^\dagger_{k,-} \end{array}\right)
\mathbf{T} (E_k/\Delta t) \sigma_3 \mathbf{T}^\dagger
\left( \begin{array}{c} a_{k,+} \\ a_{k,-} \end{array}\right) ,
\end{equation}
In the long-wavelength limit we can expand $\mathbf{T} (E_k/\Delta t) \sigma_3 \mathbf{T}^\dagger$ to first order to get
\begin{equation}
\hat{H} \approx \int dk \left( \begin{array}{cc} a^\dagger_{k,+} & a^\dagger_{k,-} \end{array}\right)
\left( pc \sigma_3 + mc^2 \sigma_2 \right)
\left( \begin{array}{c} a_{k,+} \\ a_{k,-} \end{array}\right) ,
\end{equation}
which is equivalent to the usual Dirac Hamiltonian for free fermions in one spatial dimension.

\subsection{A Dirac sea?}

A concern about the QCA construction above is that it includes negative energy states, and as $\Delta x \rightarrow 0$ the energy is unbounded below.  This is not surprising.  A similar problem arose in Dirac's original work \cite{Dirac30}.  To match the universe that we observe, it may be possible to invoke the same solution that Dirac postulated:  that the natural vacuum of the theory is not the state $\ket\Omega$ that contains no particles, but rather a vacuum $\ket{\Omega'}$ corresponding to a ``Dirac sea,'' in which all positive energy states are vacant, while all negative energy states are occupied.

The Hilbert space we considered above included only states with a finite number $N$ of particles, and therefore does not include the state $\ket{\Omega'}$.  However, if we begin with such a state $\ket{\Omega'}$, we can follow the rest of the theory above fairly closely.  The new vacuum $\ket{\Omega'}$ is the starting point.  Constructing the space of physical states is done by applying a finite set of creation operators $b^\dagger_{k,+}$ (which create particles in positive energy states) and annihilation operators $b_{k,-}$ (which destroy particles in negative energy states) to the vacuum $\ket{\Omega'}$ to construct the basis states.  If we define the energy of $\ket{\Omega'}$ to be zero, then all the physical states have positive energy.  The annihilation operators $b_{k,-}$ play the role of creation operators for antiparticles, just as in the usual Dirac field.

\section{Higher dimensions}

The above construction shows that in one spatial dimension a simple quantum cellular automaton can give rise to Dirac fields in the long-wavelength limit.  This is a very appealing result; the one-particle sector of this QCA matches the single-particle quantum walk which has already been shown to give the single-particle Dirac equation in the long-wavelength limit.  This latter result also applies in two and three spatial dimensions.  Can we construct an analogous QCA in two and three spatial dimensions that also gives rise to a Dirac field?

Though we believe that this can be done, the type of construction used in the one-dimensional QCA presented in this paper and in \cite{Arrighi20} does not generalize to higher spatial dimensions.  The problem of constructing such a QCA is closely related to the problem of {\it fermionization} in higher spatial dimensions \cite{Derzhko01}, because we must establish an ordering relation between the different fermionic modes of the theory.  We use that ordering to define a relationship between the basis states of the QCA and a set of anticommuting creation and annihilation operators.  These operators must evolve in a natural way under the QCA dynamics.  Unfortunately, as we shall see, this is very difficult to do in more than one dimension unless the underlying dynamics is {\it nonlocal}---that is, unless it is {\it not} a QCA.

An intuition for this can be seen by looking at the evolution operator in the 1D construction above.  At each time step a creation operator $a^\dagger_{x,\varepsilon}$ should evolve to a linear combination of creation operators at $x+\Delta x$ and $x-\Delta x$.  That means that two creation operators with a particular ordering, say $a^\dagger_{x,\varepsilon_1} a^\dagger_{x+\Delta x ,\varepsilon_2}$, can evolve to have the opposite ordering.  Because the creation operators anticommute, they should acquire a phase of $-1$ in that case.  In the 1D construction above they do, because when particles pass each other they acquire exactly that phase.  That is, the anticommutation relations can be enforced by a {\it local} interaction between pairs of particles.  However, while in 1D particles must pass each other to change order, in 2D or 3D the ordering may change even if two particles are physically far apart from each other.  This implies that such a phase change arising from a physical interaction would require that interaction to be nonlocal.

We now make this argument more precise.  For simplicity we will consider the case of two spatial dimensions, but the argument is exactly the same in 3D.

\subsection{Desirable properties for a QCA}

We would like to generalize the above construction for 1D, in a way analogous to how the 2D and 3D quantum walks are generalized from the 1D quantum walk in \cite{MlodinowBrun18}.  As in the 1D construction above, we would like to be able to define a set of creation and annihilation operators that are closely related to the basis states of the QCA, that obey the usual anticommutation relations, and that transform in a simple way under the evolution operator $\hat{U}$.  We want to require the following:
\begin{itemize}
\item The evolution operator takes the form
\begin{equation}
\hat{U} = \hat{C}\hat{\Sigma}_y\hat{\Sigma}_x,
\end{equation}
where $\hat{\Sigma}_x$ and $\hat{\Sigma}_y$ shift particles in the $\pm X$ and $\pm Y$ directions, depending on the internal states of the particles, and $\hat{C}$ transforms the internal states.

\item There is a set of basis states for the QCA of the form
\[
\ket{\mathbf{x}_1,\varepsilon_1; \mathbf{x}_2,\varepsilon_2; \cdots; \mathbf{x}_n,\varepsilon_n}
\]
for an $n$-particle state, where the positions are given by a vector $\mathbf{x}_i = (x_i, y_i)$, and $\varepsilon_i$ is an internal state.  We assume that the internal state takes one of a finite set of values; in 2D, this can be just the two values $\varepsilon = \pm$, while in 3D the internal state is four-dimensional.  There can be no more than one particle at a given location in a given internal state.  To avoid having multiple representations of the same basis state, we must establish an {\it ordering} convention, so we can write $\mathbf{x}_1,\varepsilon_1 <  \mathbf{x}_2,\varepsilon_2 < \cdots < \mathbf{x}_n,\varepsilon_n$.

In 1D it is easy to establish such an ordering, but it is possible in any dimension.  For example, in 2D we can say $\mathbf{x}_1, \varepsilon < \mathbf{x}_2,\varepsilon_2$ if (a) $y_1 < y_2$, or (b) $y_1=y_2$ and $x_1 < x_2$, or (c) $\mathbf{x}_1 = \mathbf{x}_2$ and $\varepsilon_1 < \varepsilon_2$ (according to some ordering of the internal states).

\item The QCA evolution is {\it local}.  Consider a given basis vector $\ket\psi$. When we apply the evolution operator $\hat{U}\ket\psi$, we get a superposition of basis vectors.  If we look at a particular location $\mathbf{x}$ on these basis vectors, the occupation of the states at that location can depend only on the occupation of that location and its neighboring sites in the original basis vector, and not on the occupation of sites far away from $\mathbf{x}$.

\item We define a set of creation operators $a^\dagger_{\mathbf{x},\varepsilon}$ which obey the usual anticommutation relations and which are related to the basis states by
\begin{equation}
\ket{\mathbf{x}_1,\varepsilon_1; \cdots; \mathbf{x}_n,\varepsilon_n} =
a^\dagger_{\mathbf{x_1},\varepsilon_1} \cdots a^\dagger_{\mathbf{x_n},\varepsilon_n} \ket\Omega .
\end{equation}
This is clearly always possible, and we get the annihilation operators as the Hermitian conjugates of the creation operators, plus the requirement that they take the vacuum state $\ket\Omega$ to 0.

\item We want theses creation and annihilation operators to evolve under $\hat{U}$ into simple linear combinations of themselves.  That is,
\begin{equation}
\hat{U} a_{(x,y),\varepsilon} \hat{U}^\dagger = \sum_{c_1=\pm 1, c_2=\pm 1, \varepsilon'}
\alpha_{c_1,c_2,\varepsilon'} a_{(x+c_1\Delta x,y+c_2\Delta x),\varepsilon'} ,
\label{eq:footprintCoeffs}
\end{equation}
where the coefficients $\alpha_{c_1,c_2,\varepsilon'}$ depend on the original internal state $\varepsilon$.  (Generically we don't expect them to depend on the position $(x,y)$, since the QCA should be translation invariant.)

\item The dynamics of the QCA are {\it nontrivial} in the following sense:  the coefficients in Eq.~(\ref{eq:footprintCoeffs}) must be nonzero for at least one internal state $\varepsilon'$ for each of the location $(x\pm\Delta x,y\pm\Delta x)$.  That is,
\[
\forall c_1,c_2 = \pm1,\  \exists \varepsilon' \ {\rm such\ that}\ \alpha_{c_1,c_2,\varepsilon'} \ne 0.
\]
\end{itemize}

We can see that in the 1D construction above, all of these conditions are satisfied.  But we will now show that in 2D (or higher dimensions), it is impossible to satisfy all of these requirements simultaneously.

\subsection{No-go theorem}

\subsubsection{Definitions}

We start by defining a few concepts that we will use in the proof.
\begin{itemize}
\item We call the combination of a particular location $\mathbf{x}$ and internal  state $\varepsilon$ a {\it site}.  A site $\mathbf{x}',\varepsilon'$ is a {\it neighbor} of the site $\mathbf{x},\varepsilon$ if $\mathbf{x} = (x,y)$ and $\mathbf{x}' = (x',y') = (x\pm\Delta x,y\pm\Delta x)$.

\item The {\it footprint} of a site $\mathbf{x},\varepsilon$ is the set of neighboring sites $(x+c_1\Delta x,y+c_2\Delta x),\varepsilon'$ such that $\alpha_{c_1,c_2,\varepsilon'} \ne 0$ for $c_1,c_2 = \pm1$.

\item Two sites $\mathbf{x}_1,\varepsilon_1$ and $\mathbf{x}_2,\varepsilon_2$ are {\it connected} if $\mathbf{x}_2,\varepsilon_2$ lies in the footprint, or the footprint of the footprint, or the footprint of the footprint of the footprint...etc., of $\mathbf{x}_1,\varepsilon_1$, or vice versa.  We call a chain of sites like this a {\it path}, where each site in the path is in the footprint of the previous site.

Note that for two sites $\mathbf{x}_1,\varepsilon_1$ and $\mathbf{x}_2,\varepsilon_2$ to be connected, if we write $\mathbf{x}_1 = (x_1,y_1) = (i_1\Delta x,j_1\Delta x)$ and $\mathbf{x}_2 = (x_2,y_2) = (i_2\Delta x,j_2\Delta x)$, then $i_1+j_1$ and $i_3+j_3$ must either be both odd or both even.  (This is a necessary condition, but for some nongeneric cases may not be sufficient).
\end{itemize}

\subsubsection{Contradiction}

Choose three operators $a_{\mathbf{x}_1,\varepsilon_1}$, $a_{\mathbf{x}_2,\varepsilon_2}$, $a_{\mathbf{x}_3,\varepsilon_3}$, where $\mathbf{x}_k = (x_k,y_k) = (i_k\Delta x,j_k\Delta x)$, which satisfy the following conditions:
\begin{enumerate}
\item the locations $\mathbf{x}_1$ and $\mathbf{x}_3$ are both far from $\mathbf{x}_2$;

\item in the canonical ordering of the basis states, $\mathbf{x}_1,\varepsilon_1 < \mathbf{x}_2,\varepsilon_2 < \mathbf{x}_3,\varepsilon_3$;

\item the sites $\mathbf{x}_1,\varepsilon_1$ and $\mathbf{x}_3,\varepsilon_3$ are connected by a path that is everywhere far from $\mathbf{x}_2,\varepsilon_2$.
\end{enumerate}

We will use these conditions to establish a contradiction, as follows.  Consider the sites in the footprint of $\mathbf{x}_2,\varepsilon_2$.  If any of these sites is $<$ any of the sites in the footprint of $\mathbf{x}_1,\varepsilon_1$, then we immediately see that $\hat{U}$ will map a basis vector where $\mathbf{x}_1,\varepsilon_1$ and $\mathbf{x}_2,\varepsilon_2$ are both occupied to a superposition of basis states, at least one term of which will acquire an extra phase of $-1$ because of the reordering of the anticommuting creation operators $a^\dagger_{\mathbf{x}_1,\varepsilon_1}$ and $a^\dagger_{\mathbf{x}_2,\varepsilon_2}$.  This phase depends on the presence of a particle far away, which contradicts the assumption that the QCA evolution is local.

Similarly, if any of the sites in the footprint of $a_{\mathbf{x}_2,\varepsilon_2}$ is  $>$ any of the sites in the footprint of $\mathbf{x}_3,\varepsilon_3$, then $\hat{U}$ will map a basis vector where $\mathbf{x}_2,\varepsilon_2$ and $\mathbf{x}_3,\varepsilon_3$ are both occupied to a superposition of basis states, at least one term of which will acquire an extra phase of $-1$ , again contradicting the assumption that the QCA evolution is local.

Suppose, then that the ordering of sites in the footprints of $\mathbf{x}_2,\varepsilon_2$ and $\mathbf{x}_1,\varepsilon_2$ all retain the same ordering of the original pair of sites (and similarly with $\mathbf{x}_3,\varepsilon_3$).  Then compare the ordering of the sites in the footprint of $\mathbf{x}_2,\varepsilon_2$ to those in the footprint of the next site along the path from $\mathbf{x}_1,\varepsilon_1$ to $\mathbf{x}_3,\varepsilon_3$.  Because the sites at the endpoints of this path have the opposite ordering with respect to $\mathbf{x}_2,\varepsilon_2$, at some point along this path there must be a site where the ordering is flipped.  By assumption 3 above, all sites along this path are far from $\mathbf{x}_2,\varepsilon_2$.  So we again contradict the locality of the QCA evolution.

This proves that a QCA construction satisfying all the desired properties above is impossible in 2D.  And it is clear that an essentially identical argument will hold in 3D or higher spatial dimensions.

\subsubsection{Loopholes}

Is there any way around this conclusion?  The trickiest assumption is assumption 3 above, that we can always find two sites $\mathbf{x}_1,\varepsilon_1$ and $\mathbf{x}_3,\varepsilon_3$ that have opposite ordering with respect to site $\mathbf{x}_2,\varepsilon_2$, and which are connected by a path that is everywhere far from $\mathbf{x}_2,\varepsilon_2$.  Our method above for deriving Dirac field theory from a QCA was successful because in 1D this assumption fails.  In 1D, if the ordering of the sites largely corresponds to their physical ordering, it is impossible to form a path that is everywhere far from $\mathbf{x}_2,\varepsilon_2$.  And indeed, it is exactly this property of the 1D case that we exploit in the construction earlier in this paper, where we add a {\it local} phase to the evolution for particles passing each other.

In 2D or 3D we could make it impossible to find two such connected sites generically by making the dynamics {\it trivial}: if all particles move in the same direction, or there is no amplitude for a particle moving in one direction to change direction, then one could establish an ordering that is never changed by the evolution $\hat{U}$.    But this contradicts the assumption of nontrivial dynamics.

Finally, if we let the internal space be very high-dimensional (so there is a very large number of possible internal states $\varepsilon$, then it might be that most pairs of sites are not connected, and the argument is no longer obviously true.

But generically, a system with nontrivial dynamics and a low-dimensional internal space will have pairs of sites that satisfy assumption 3 above, and hence contradict the existence of a QCA with our set of desired properties in two dimensions or higher.

\section{Discussion}

It has been shown that a quantum walk with simple properties leads, in the long-wavelength limit, to the free Dirac equation in 3+1 dimensions.  Our aim is eventually to show that there exists an analogous QCA that reproduces Dirac field theory in 3+1 dimensions, along with interactions---i.e., QED.  Here we have taken the first step by showing that in 1D it is possible to construct a QCA which corresponds to a field theory for free fermions in the long wavelength limit.  However, we have also shown that this type of construction cannot be generalized to two or three spatial dimensions; a local QCA with nontrivial dynamics does not generically allow the existence of creation and annihilation operators that act at local sites, that obey the usual anticommutation relations, and that evolve under the QCA dynamics into simple linear combinations of themselves.

Does this mean that it is impossible to find a QCA model that give a field theory for free fermions in the long wavelength limits?  We believe not.  We believe it is possible to weaken the assumptions of the no-go theorem to the point where it no longer holds.  Several possibilities suggest themselves.  Perhaps the anticommutation relations themselves only hold approximately in the long-wavelength limit.  Perhaps one can find a set of creation and annihilation operators that do not act locally (in the sense assumed in this paper), but rather are ``smeared out'' in space.  That is true in the usual continuous field theories; it might well be true for QCA models as well.  Or perhaps the assumption of a low-dimensional internal space is flawed.  Perhaps the QCA model has local subsystems that are high-dimensional, and the usual low-dimensional theory only arises again in the long-wavelength limit.  Or some other strategy for increasing the dimension of the Hilbert space might get around this result.  Given the striking success of single-particle quantum walk theories, we are now considering these possibilities.  Success would fulfill Feynman and Wheeler's vision of the material universe as a giant quantum computer.

\begin{acknowledgments}

TAB acknowledges useful conversations with Namit Anand, Christopher Cantwell, Yi-Hsiang Chen, Shengshi Pang, Prithviraj Prabhu and Chris Sutherland.  LM would like to thank Erhard Seiler and Alois Kabelschacht of the Max-Planck-Institute in Munich for fruitful discussions.  The authors are grateful for the hospitality of John Preskill and Caltech's Institute for Quantum Information and Matter (IQIM).

\end{acknowledgments}


\begin{thebibliography}{99}

\bibitem{Wheeler89} J.A. Wheeler, {\sl Information, Physics, Quantum: The Search for Links}, Santa Fe Institute Conferences, 29 May--2 June and 4-8 June 1989.

\bibitem{Feynman82} R.P. Feynman, {\sl Simulating Physics With Computers}, collected in {\sl Feynman And Computation}, A.J.G. Hey (ed.) (Perseus Books, Cambridge, MA, 1999).

\bibitem{BrunMlodinow19} Todd A. Brun and L. Mlodinow, {\sl Detection of discrete spacetime by matter interferometry}, Phys. Rev. D {\bf 99}, 015012 (2019).

\bibitem{AharonovY93} Y. Aharonov, L. Davidovich, and N. Zagury, {\sl Quantum random walks}, Phys. Rev. A {\bf 48}, 1687 (1993).

\bibitem{Ambainis01} A. Ambainis, E. Bach, A. Nayak, A. Vishwanath and J. Watrous, {\sl One-dimensional quantum walks}, in {\sl Proceedings of the ACM Symposium on Theory of Computation (STOC `01), 2001} (Association for Computing Machinery, New York, 2001), pp. 37--49.

\bibitem{AharonovD01} D. Aharonov, A. Ambainis, J. Kempe and U. Vazirani, {\sl Quantum Walks On Graphs}, in {\sl Proceedings of the ACM Symposium on Theory of Computation (STOC `01), 2001} (Association for Computing Machinery, New York, 2001), pp. 50--59.

\bibitem{Kempe03} J. Kempe, {\sl Quantum walks---an introductory overview}, Contemporary Physics {\bf 44}, 307--327 (2003).

\bibitem{Meyer96a} D. A. Meyer, {\sl  On the absence of homogeneous scalar unitary cellular automata}, Phys. Lett. A {\bf 223}, 337--340 (1996).

\bibitem{Moore02} C. Moore and A. Russell, {\sl Quantum walks on the hypercube} in {\sl Proceedings of the 6th International Workshop on Randomization and Approximation Techniques}, 164--178 (Springer-Verlag, London, 2002).

\bibitem{Yin08} Y. Yin, D. E. Katsanos, and S. N. Evangelou, {\sl Quantum walks on a random environment}, Phys. Rev. A {\bf 77}, 022302 (2008).

\bibitem{Brun03a} T.A. Brun, H.A. Carteret and A. Ambainis, {\sl The quantum to classical transition for random walks}, Phys. Rev. Lett. {\bf 91}, 130602 (2003).

\bibitem{Brun03b} T.A. Brun, H.A. Carteret and A. Ambainis, {\sl Quantum random walks with decoherent coins}, Phys. Rev. A {\bf 67}, 032304 (2003).

\bibitem{Kendon03} V. Kendon and B. Tregenna, {\sl Decoherence can be useful in quantum walks}, Phys. Rev. A {\bf 67}, 042315 (2003).

\bibitem{Brun03c} T.A. Brun, H.A. Carteret and A. Ambainis, {\sl Quantum Walks driven by many coins}, Phys. Rev. A {\bf 67}, 052317 (2003).

\bibitem{Kendon06} V. Kendon, {\sl Decoherence in quantum walks---a review}, Math. Struct. in Comp. Sci {\bf 17}, 1169--1220 (2006).

\bibitem{Childs02} A. M. Childs, E. Farhi, and S. Gutmann, {\sl An example of the difference between quantum and classical random walks}, Quant. Inform. Proc. {\bf 1}, 35 (2002).

\bibitem{Shenvi03} N. Shenvi, J. Kempe, and K. Birgitta Whaley, {\sl Quantum random-walk search algorithm}, Phys. Rev. A {\bf 67}, 052307 (2003).

\bibitem{Ambainis03} A. Ambainis, {\sl Quantum walks and their algorithmic applications}, Int. J. Quant. Inf. {\bf 1}, 507 (2003).

\bibitem{Ambainis07} A. Ambainis, {\sl Quantum walk algorithm for element distinctness}, SIAM J. Comput. {\bf 37}, 210 (2007).

\bibitem{Farhi08} E. Farhi, J. Goldstone, and S. Gutmann, {\sl A quantum algorithm for the Hamiltonian NAND tree}, Theo. Comput. {\bf 4}, 169 (2008).

\bibitem{Ambainis10} A. Ambainis, A. M. Childs, B. W. Reichardt, R. \v{S}palek and S. Zhang, {\sl Any AND-OR Formula of Size N Can Be Evaluated in Time $N^{1/2+o(1)}$ on a Quantum Computer}, SIAM J. Comput. {\bf 39}, 2513--2530 (2010).

\bibitem{Reichardt12} B. W. Reichardt and R. \v{S}palek, {\sl Span-program-based quantum algorithm for evaluating formulas}, Theo. Comput. {\bf 8}, 291 (2012).

\bibitem{Bialynicki94} I. Bialynicki-Birula, {\sl Weyl, Dirac, and Maxwell equations on a lattice as unitary cellular automata}, Phys. Rev. D {\bf 49}, 6920 (1994).

\bibitem{Meyer96} D.A. Meyer, {\sl From quantum cellular automata to quantum lattice gases}, J. Stat. Phys. {\bf 85}, 551--574 (1996).

\bibitem{Strauch06} F. W. Strauch, {\sl Connecting the discrete- and continuous-time quantum walks}, Phys. Rev. A {\bf 74}, 030301(R) (2006).

\bibitem{Bracken07} A. J. Bracken, D. Ellinas and I. Smyrnakis, {\sl Free-Dirac-particle evolution as a quantum random walk}, Phys. Rev. A {\bf 75}, 022322 (2007).

\bibitem{Chandrashekar10} C. M. Chandrashekar, S. Banerjee and R. Srikanth, {\sl Relationship between quantum walks and relativistic quantum mechanics}, Phys. Rev. A {\bf 81}, 062340 (2010).

\bibitem{Chandrashekar11} C. M. Chandrashekar, {\sl Two-state quantum walk on two- and three-dimensional lattices}, arXiv:1103.2704.

\bibitem{DAriano14} G. M. D'Ariano and P. Perinotti, {\sl Derivation of the Dirac equation from principles of information processing}, Phys. Rev. A {\bf 90}, 062106 (2014).

\bibitem{Arrighi14} P. Arrighi, M. Forets and Vincent Nesme, {\sl The Dirac equation as a quantum walk: higher dimensions, observational convergence}, J. Phys. A {\bf 47}, 465302 (2014).

\bibitem{Chandrashekar13} C. M. Chandrashekar, {\sl Two-component Dirac-like Hamiltonian for generating quantum walk on one-, two- and three-dimensional lattices}, Scientific Reports {\bf 3}, 2829 (2013).

\bibitem{Farrelly14} T. C. Farrelly and A. J. Short, {\sl Discrete spacetime and relativistic quantum particles}, Phys. Rev. A {\bf 89}, 062109 (2014).

\bibitem{DAriano15} G. M. D'Ariano, N. Mosco, P. Perinotti and A. Tosini, {\sl Discrete Feynman propagator for the Weyl quantum walk in 2+1 dimensions}, Europhy. Lett. {\bf 109}, 40012 (2015).

\bibitem{Succi15} S. Succi, F. Fillion-Gourdeau and S. Palpacelli, {\sl Quantum Lattice Boltzmann is a quantum walk}, EPJ Quantum Technology {\bf 2}, 12 (2015).

\bibitem{Bisio15} Alessandro Bisio, Giacomo Mauro D'Ariano and Alessandro Tosini, {\sl Quantum Field as a quantum cellular automaton: the Dirac free evolution in one dimension}, Annals of Physics {\bf 354}, 244--264 (2015).

\bibitem{Arrighi15} P. Arrighi, S. Facchini and M. Forets, {\sl Quantum walking in curved spacetime}, arXiv:1505.07023.

\bibitem{Bisio16} Alessandro Bisio, Giacomo Mauro D'Ariano, Marco Erba, Paolo Perinotti, Alessandro Tosini, {\sl Quantum walks with a one-dimensional coin}, Phys. Rev. A {\bf 93}, 062334 (2016).

\bibitem{Arrighi16} P. Arrighi and S. Facchini, {\sl Quantum walking in curved spacetime:  $(3+1)$ dimensions, and beyond}, arXiv:1609.00305.

\bibitem{DAriano17} G. M. D'Ariano, {\sl Physics Without Physics: the Power of Information-theoretical Principles}, Int. J. Theor. Phys. {\bf 56}, 97 (2017).

\bibitem{Raynal17} Phillippe Raynal, {\sl Simple derivation of the Weyl and Dirac quantum cellular automata}, Phys. Rev. A {\bf  95}, 062344 (2017).

\bibitem{DAriano17b} G. M. D'Ariano, M. Erba and P. Perinotti, {\sl Isotropic quantum walks on lattices and the Weyl equation}, Phys. Rev. A {\bf 96}, 062101 (2017).

\bibitem{MlodinowBrun18} L. Mlodinow and T.A. Brun, {\sl Discrete spacetime, quantum walks and relativistic wave equations}, Phys. Rev. A {\bf 97}, 042131 (2018).

\bibitem{Bisio18} A. Bisio, G.M. D’Ariano, N. Mosco, P. Perinotti and A. Tosini, {\sl Solutions of a Two-Particle Interacting Quantum Walk}, Entropy {\bf 20}, 435 (2018).

\bibitem{DAriano19} G.M. D'Ariano, M. Erba and P. Perinotti, {\sl Chirality from quantum walks without a quantum coin}, Phys. Rev. A {\bf 100}, 012105 (2019).

\bibitem{Maeda20} M. Maeda and A. Suzuki, {\sl Continuous limits of linear and nonlinear quantum walks}, Rev. Math. Phys. {\bf 32}, 2050008 (2020).

\bibitem{Manighalam19} M. Manighalam and M. Kon, {\sl Continuum Limits of the 1D Discrete Time Quantum Walk}, arXiv:1909.07531 (2019).

\bibitem{Arrighi14b} P. Arrighi, S. Facchini and M. Forets, {\sl Discrete Lorentz covariance for Quantum Walks and Quantum Cellular Automata}, New J. Phys. {\bf 16}, 093007 (2014).

\bibitem{Bisio17} A. Bisio, G.M. D'Ariano, P. Perinotti, {\sl Quantum Walks, Weyl equation and the Lorentz group}, Found. Phys. {\bf 47}, 1065--1076 (2017).

\bibitem{BjorkenDrell65} J.D. Bjorken and S.D. Drell, {\sl Relativistic Quantum Fields} (McGraw-Hill, New York, 1965).

\bibitem{Watrous91} J. Watrous, {\sl On one-dimensional quantum cellular automata}, Complex Systems {\bf 5}, 19 (1991).

\bibitem{Lloyd93} S. Lloyd, {\sl A Potentially Realizable Quantum Computer}, Science {\bf 261}, 1569--1571 (1993).

\bibitem{Watrous95} J. Watrous, {\sl On one-dimensional quantum cellular automata}, in {\sl Proceedings of the 36th Annual Symposium on Foundations of Computer Science}, 528--537 (1995). 

\bibitem{Grossing88} G. Gr\"ossing and A. Zeilinger, {\sl Quantum cellular automata}, Complex Systems {\bf 2}, 197--208 (1988).

\bibitem{Fussy93} S. Fussy, G. Gr\"ossing, H. Schwabl and A. Scrinzi, {\sl Nonlocal computation in quantum cellular automata}, Phys. Rev. A {\bf 48}, 3470--3477 (1993).

\bibitem{Durr96} C. Durr and M. Santha, {\sl A decision procedure for unitary linear quantum cellular automata}, in {\sl Proceeding of the 37th IEEE Symposium on Foundations of Computer Science}, 38--45 (1996).

\bibitem{Meyer96b} D.A. Meyer, {\sl Unitarity in one dimensional nonlinear quantum cellular automata}, arXiv:quant-ph/9605023.

\bibitem{Bisio15b} A. Bisio, G.M. D’Ariano, P. Perinotti and A. Tosini, {\sl Free quantum field theory from quantum cellular automata}, Found. Phys. {\bf 45}, 1137--1152 (2015).

\bibitem{Bisio15c} A. Bisio, G.M. D’Ariano, P. Perinotti and A. Tosini, {\sl Weyl, Dirac and Maxwell quantum cellular automata}, Found. Phys. {\bf 45}, 1203--1221 (2015).

\bibitem{Bisio16b} A. Bisio, G.M. D’Ariano and P. Perinotti, {\sl Quantum Cellular Automaton Theory of Light}, Ann. Phys. {\bf 368}, 177--190 (2016).

\bibitem{Mallick16} A. Mallick and C.M. Chandrashekar, {\sl Dirac Cellular Automaton from Split-step Quantum Walk}, Sci. Reports {\bf 6}, 25779 (2016).


\bibitem{Farrelly19} T. Farrelly, {\sl A Review of Quantum Cellular Automata}, arXiv:1904.13318.

\bibitem{Shakeel13} A. Shakeel and P.J. Love, {\sl When is a Quantum Cellular Automaton (QCA) a Quantum Lattice Gas (QLGA)?}, J. Math. Phys. {\bf 54} 092203 (2013).

\bibitem{Yepez13} J. Yepez, {\sl Quantum Lattice Gas Model of Dirac Particles in 1+1 Dimensions}, arXiv:1307.3595.

\bibitem{Yepez16} J. Yepez, {\sl Quantum Lattice Gas Algorithmic Representation of Gauge Field Theory}, Proc. SPIE 9996, {\sl Quantum Information Science and Technology II}, 99960N (24 October 2016).




\bibitem{Jordan28} P. Jordan and E. Wigner, {\sl Uber das Paulische Aquivalenzverbot}, Z. Phys. {\bf 47}, 631 (1928).

\bibitem{Arrighi20} P. Arrighi, C. B\'eny and T. Farrelly, {\sl A quantum cellular automaton for one-dimensional QED}, Quant. Inf. Proc. {\bf 19}, 88 (2020).

\bibitem{Dirac30} P.A.M. Dirac, {\sl  A Theory of Electrons and Protons}, Proc. R. Soc. Lond. A, {\bf 126}, 360--365 (1930).

\bibitem{Derzhko01} O. Derzhko, {\sl Jordan-Wigner Fermionization for spin-1/2 systems in two dimensions:  a brief review}, J. Physics Studies {\bf 5}, 49 (2001).

\end{thebibliography}

\end{document}